\documentclass[12pt, preprint]{aastex}
\usepackage{emulateapj5}  

\def\ergcm2s{~erg cm$^{-2}$ s$^{-1}$ } 
\def\ergs{~erg s$^{-1}$}		
\def\cmsq{~cm$^{-2}$ }		%
\def\nh{$\rm{N_{H}}$}

\def\etal{et al.~}		

\def\msunpyr{~M$_{\odot}~\rm{yr^{-1}}$}
\def\deg{$^{\circ}$}
\def\n4038{~NGC4038/39}		
\def\chandra{{\it Chandra }}

\def\x2{$\chi^{2}$}	



\begin{document}

\title{\chandra observations of the X-ray luminous star-forming galaxy 
merger Arp~299\\} 

\author{A. Zezas}
\affil{Harvard-Smithsonian Center for Astrophysics, 60 Garden
Street, Cambridge, MA 02138; azezas@cfa.harvard.edu }

\author{M.J. Ward}
\affil{X-ray Astronomy Group, Leicester University, University Road,
LE2 7RH, Leicester, UK;  mjw@star.le.ac.uk}

\author{S.S. Murray}
\affil{Harvard-Smithsonian Center for Astrophysics, 60 Garden
Street, Cambridge, MA 02138; smurray@cfa.harvard.edu}

\shorttitle{\chandra\ observations of Arp~299}
\shortauthors{Zezas et al}

\begin{abstract}
We report results of a {\it Chandra} observation of the
X-ray luminous star-forming galaxy Arp299 (NGC3690/IC694). We detect
18 discrete X-ray sources with luminosities above
$\sim10^{39}$\ergs~(0.5-8.0~keV band), which
contribute $\sim40\%$ of the total galactic emission in this 
band. The remaining emission is associated with a diffuse component
spatially coincident with regions of widespread star-formation. We
detect X-ray emission from both nuclei. One of the discrete sources within the 
complex nuclear region of NGC~3690 is found to have a
very hard spectrum and therefore we associate it with the origin of the AGN-like
spectrum that has also been detected at high X-ray energies using Beppo-SAX.    
\end{abstract}


 \section{Introduction}

Arp299 is one of the nearest (D=40.4~Mpc\footnote{Throughout this
paper we assume an $\rm{H_{o}=75~km/s/Mpc}$ and $\rm{q_{o}=0.0}$})
Luminous Infrared
Galaxies (LIRGs; Sanders \& Mirabel 1996; $\rm{L_{IR}=5\times10^{11}~L_{\odot}}$) 
and is therefore an excellent 
candidate for detailed studies of their energy source\footnote{In some 
papers Arp299 is referred to as a ULIRG, based on the use of $\rm{H_{o}=50~km/s/Mpc}$}. Optical, 
infrared (IR) and radio observations show that it hosts galaxy-wide
star-formation with estimated rates for discrete regions ranging from  $\sim20$\msunpyr~ up to
$\sim140$\msunpyr~  and typical ages of $\sim10$~Myr (Alonso-Herrero
\etal 2000 henceforth referred to as AH00; Zhao \etal 1997). 
Mid IR observations indicate that the major part of
this star-forming activity is obscured by large amounts of dust
(Charmandaris \etal 2002). 
 
X-ray observations show that Arp299 is one of
the most X-ray luminous star-forming galaxies in the  
nearby universe, with an observed luminosity of
$4\times10^{41}$~\ergs~ in the 0.1-10.0~keV band (Zezas \etal 1998;
Heckman \etal 1999). Its X-ray spectrum is well
represented by a two component model with either a double  thermal
plasma or a thermal plasma and a power-law, and was found to be very
similar to that of other 
much less luminous star-forming galaxies. However, recent observations of Arp~299 with Beppo-SAX 
revealed a heavily obscured AGN (Della Ceca \etal 2002). In order to further study the
origin of the X-ray emission from this exceptionally luminous
star-forming galaxy  we obtained a relatively short exposure with the
Chandra X-ray observatory  (Weisskopf \etal 2000).  In this paper we report on the properties
of the 
discrete sources in Arp~299. A more detailed study of the X-ray emission from this system
will be presented in a forthcoming paper.

\section{Observations and Data Analysis}

Arp~299 was observed with the ACIS-I camera (Garmire \etal 1997) on
board \chandra for 24.2~ks on the 13th of July 2001.  In our analysis
we used the event-2 type files and followed the standard procedures
for analysis of \chandra data\footnote{http://asc.harvard.edu} using
the CIAO v2.2 tool suite.  We corrected the absolute astrometry for
known
offsets\footnote{http://asc.harvard.edu/ciao/threads/arcsec\_correction/}. Comparison
between the positions of the brightest isolated sources in the field
and the coordinates of their radio counterparts (kindly provided by
S. Neff) showed that the absolute astrometry is accurate to within
$1''$.  The background level during our observation was nominal.

We created images in the full (0.5-8.0~keV) as well as soft
(0.5-2.5~keV), medium (2.5-4.0~keV) and hard (4.0-8.0~keV) bands.  We
adaptively smoothed these images using the {\textit{csmooth}} tool,
with a maximum signal-to-noise (S/N) ratio of 5. In order to avoid
over-smoothing weak features we set the maximum FWHM of the smoothing
Gaussian to 16$''$, resulting in a minimum S/N of $\sim2$ in regions
of the galaxy with weak X-ray emission.  Figure~1 shows a true colour X-ray image of
the galaxy, constructed from the combined soft (red), medium (green)
and hard (blue) band smoothed images.  It is clear from this image
that a significant fraction of the X-ray emission from the galaxy is
associated with a diffuse soft component and that there are also a few hard
discrete sources.

We searched the images in each band for discrete sources  using the
CIAO \textit{wavedetect} algorithm (Freeman \etal 2002).  This
resulted in the detection  of 18 discrete sources, 4 of
which appear to have an extended component based on the
{\textit{wavdetect}} results. The properties of these sources are
presented in Table~1.  Figure~2 shows a full band image
with the discrete sources indicated following the notation of Table~1.
We estimated the intensity of each source from within an aperture
including most of the observed emission (in the case of point-like
sources the aperture radius was $\sim1.5''$ and encompassed 90\% of the total
emission at 1.4~keV).  These apertures are shown projected onto the X-ray image in
Fig.~2.  The
background was derived from an annulus placed around each source (excluding
any encompassed sources).  The absorption corrected luminosity of the
sources range from $\sim10^{39}$\ergs~ up to $2.4\times10^{40}$~\ergs~
(0.1-10.0~keV), assuming a power-law ($\Gamma=1.7$) with the Galactic
line-of-sight HI column density (\nh=$1.0\times10^{20}$\cmsq; Stark
\etal 1992). We note that because of the varying intensity of the  diffuse
emission over the galaxy, the detection limit is not uniform. The
discrete sources contribute $\sim40$\% of the galactic X-ray emission
in the broad 0.5-8.0~keV energy band. Their contributions are $\sim35$\% and
65\% in the soft and medium bands, respectively.

For the 6 X-ray sources with more than 80 counts,
(ie. Src2 = C1 in NGC~3690; Src3; Src4 = B2 in NGC~3690;
Src6 = B1 in NGC~3690; Src14 and Src16 = A in IC~694, following the notation of AH00),  we extracted X-ray spectra 
from the above apertures. We fitted these spectra using the XSPEC
v11.2 spectral fitting package. In order to take account of the
absorption due to the build-up on the window of the ACIS detector we
used the {\textit{acisabs}}
model{\footnote{http://asc.harvard.edu/cal/Acis/Cal\_prods/qeDeg/ \\
http://www.astro.psu.edu/users/chartas/xcontdir/xcont.html}}.  The
results of the spectral fits are presented in Table~2.
In all but one case (Src6) a power-law or a Raymond-Smith 
thermal plasma with an absorption column left free to vary, gave good fits.
For these sources we derived  power-law photon indices between $\Gamma=1.4$
and $\Gamma\sim2.7$. The best fit column density was consistent with values significantly in
excess of the Galactic line of sight value in all cases. 

For source Src~6, in order to obtain a reasonable fit an additional
spectral component was required.  We obtained the best fit ($\rm{\chi^{2}_{\nu}}=8.3/9$) using a
double power-law model (note: a thermal plasma plus absorbed power-law
model did not give a good fit; $\rm{\chi^{2}_{\nu}}=12.2/8$). In this case the best fit photon
indices were $\Gamma_{1}=-2.46$ and $\Gamma_{2}=1.89$, while the
overall absorption was moderate (\nh=$1.9\times10^{21}$\cmsq). By
introducing a second absorber for one of the power-law components we
obtain a photon index $\Gamma=1.6$ and a column density for the second
power-law component of ~\nh=$8.9\times10^{23}$\cmsq and a fit of
similar quality ($\rm{\chi^{2}_{\nu}}=7.7/9$; the indices of
the two power-laws were bound together, since when fitted individually they were poorly
constrained).
Our results strongly suggest
that the AGN in Arp299 recently revealed at high X-ray energies (Della
Ceca \etal 2002), is associated with this source, identified with the infrared source 
B1. The high spatial resolution of \chandra allows the source to
be separated from the surrounding emission, and hence permits a much
improved study of the spectrum below 10~keV than was possible using
the Beppo-SAX data.  We also modeled the spectrum with the best fit
spectrum reported in Dela Cecca \etal (2002),
keeping the parameters of the absorbed component (slope and
normalization) fixed to the values determined from the Beppo-SAX
data while leaving the other parameters  free
to vary. Our results show that neither a thermal component nor a
strong Fe-K$\alpha$ line are required by the fit: the former most
probably because of the very small region used in the spectral
extraction, while the latter was because of the poor photon statistics in the
6~keV region. This fit gave an unabsorbed power-law of slope
$\Gamma=1.26^{+0.3}_{-0.14}$, an overall ~\nh~ of
$6.7\times10^{20}$\cmsq ($<2.3\times10^{21}$\cmsq) and a column
density for the hard component of
$2.5_{-0.25}^{+0.44}\times10^{24}$\cmsq, very similar to the values derived from the
Beppo-SAX data.
For the other sources we calculated hardness ratios (Table~1), which
are consistent with power-laws in the range $\Gamma= 1.5-3.0$.

\subsection{Multiwavelength associations} 

We have compared the X-ray data with observations made in the radio
(Zhao \etal 1997; Neff \etal private communication, 2002) and at NIR
wavelengths (AH00).  From complex-A we detect X-ray emission
associated with sources A (=IC~694), A3-4 and A6 (following the
notation of AH00).  From complex-B we detect the X-ray counterparts of
components B1 (=NGC~3690), B2, B21 and possibly B23. Finally from
complex-C we detect a discrete X-ray source coincident with components
C1-2 as well as diffuse emission in the neighborhood of C4 and C5.

In the radio regime, we can identify X-ray counterparts for all of the
major radio components (A, B1, B2, C) based on the accurate
radio coordinates kindly provided by S. Neff (private communication).
We also find an X-ray counterpart (Src~8)  for a radio source in the
East of C$'$, which we will refer to as C$''$
($\alpha=11^{h}28^{m}31.7^{s}$, $\delta=+58^{\circ}33^{'}49.5^{''}$; J2000).
In addition we have detected an X-ray counterpart (Src10) of the radio source
IC~694+1990 (which lies 5$''$ south-west of component A (Huang \etal
1990; Zhao \etal 1997; Neff \etal 2002) which is also associated with
a NIR source (not identified by AH00).
Its relatively soft X-ray colors suggest that it may be associated
with a radio supernova, as was originally proposed by Huang
\etal (1990), rather than a background QSO (Zhao \etal 1997). 

\section{Discussion}

\subsection{Discrete sources}

This relatively short duration \chandra observation of Arp299 shows
that 40\% of the galactic X-ray emission in the (0.5-8.0)~keV band
arises from 18 discrete sources with luminosities above
$\sim10^{39}$~\ergs, which is similar to results obtained for other
star-forming galaxies eg. the Antennae 
(Fabbiano \etal 2001), M~82 (Zezas \etal 2001) and NGC~3256 (Lira
\etal 2002).
All of the major optical, NIR and radio components of 
the system (including the two nuclei) are detected in X-rays in this \chandra observation.  

We have compared the luminosity distribution of the point-like sources
in Arp~299 with that for sources detected in NGC~3256 (D=34~Mpc; Lira
\etal 2002) and the Antennae (Zezas \& Fabbiano 2002), after rebinning
the X-ray data for the latter by a factor of 2 in order to match the
spatial resolution of Arp~299.  
We find that the luminosity distributions, excluding the extended sources and the
nuclei, are very similar indicating that these galaxies have comparable
intrinsic numbers of luminous sources.
Therefore the apparently large number of sources with luminosities
above $10^{39}$~\ergs~ in both Arp~299 and NGC~3256 compared with the Antennae,
is most likely due to source confusion 
(at 40~Mpc one resolution element corresponds to a physical scale of
$\sim100$~pc).

\subsection{The nuclei}

The excellent spatial resolution of \chandra allows us to individually
study the two nuclei and to compare their X-ray properties with those
at other wavelengths.  The X-ray spectrum of Source~16
which is associated with the nucleus of IC~694 (infrared source A) is well fit with a
heavily absorbed (\nh$\sim1.2\times10^{22}$\cmsq) power-law
($\Gamma\sim1.4$) consistent with either a population of X-ray
binaries or an AGN. Its observed luminosity in the 0.1-10.0~keV band
is $\sim3.9\times10^{39}$\ergs, that is $\sim5$\% of the overall
emission from Arp~299.  
The measured absorbing column density is
consistent with the extinction derived from mid-IR observations
($\rm{A_{V}=17}$ mags; Charmandaris \etal 2002), and narrow-band NIR
photometry and spectroscopy (AH00). 

Although the contribution of the nuclear point
source (Src~16) to the overall X-ray emission of Arp~299 is small, we detect
a significant circumnuclear extended component
($\sim3.4\times1.7$~kpc). 
This region has an absorption corrected luminosity of
$3\times10^{40}$\ergs~ (0.1-10.0~keV; providing 8\% of the observed
 emission Arp~299 in this band). 
Based on modeling of the stellar population, the dominant stellar component in
nucleus A has an age of 11~Myr.
Given the large young stellar population, this region may be expected to contain
a significant number of High Mass X-ray Binaries (HMXB).  Taking the
derived Lyman continuum flux (Ly$_c$) of this region (Zhao \etal 1997;
AH00) and assuming a mean Ly$_c$ flux of $\sim10^{49}~\rm{s^{-1}}$ per
O-type star, we estimate a population of $\sim 5\times10^{5}$ O-type
stars.  Based on the minimum specific X-ray luminosity per O-type star
found in the SMC (Helfand \& Moran 2001) we estimate a luminosity of
$3.5\times10^{40}$~\ergs~ for the HMXB component, which is in good agreement with the
intrinsic X-ray luminosity from this region.

Turning now to the case of X-ray Source~6, which we identify with the
nuclear component B1 in NGC~3690, we find that it has a very
hard X-ray spectrum typical of a heavily obscured AGN, indicating that 
this component is the nucleus of NGC~3690.  In the light of
these spectroscopic results we associate the recently discovered
highly obscured AGN in Arp~299 (Della-Ceca \etal 2002) with the
nucleus of NGC~3690.  The observed luminosity of this nucleus is
$\sim6.3\times10^{40}$~\ergs ~(0.1-10~keV), and contributes 2\% and
15\% of the overall X-ray emission of Arp~299 in the 0.1-2.0~keV and
2.0-10.0~keV X-ray bands respectively. However, as reported by Della
Ceca \etal (2002), even  its absorption corrected intrinsic hard X-ray luminosity,
is only a small fraction of the bolometric energy output of
Arp~299. From the combined Beppo-SAX \chandra fits we find that the
spectrum below 10.0~keV is dominated by a relatively flat power-law,
which could either be associated with scattered emission from the AGN
or with a population of HMXBs in the circumnuclear region. The
intrinsic luminosity of this spectral component is $\sim3.0\times10^{40}$\ergs~
in the 0.1-10.0keV band, which is relatively high for a population of
X-ray binaries given the physical size of the extraction region,
unless it is dominated by a few Ultra-Luminous X-ray Sources (ULXs) as has been
seen in M~82 (e.g. Kaaret \etal 2001). Given the presence of
diffuse  hard X-ray emission (extending up to 8~keV) in the region of NGC~3690, some fraction
of this luminosity might be associated with a starburst
component. However, from Fig.~1, we do see a clear displacement
between the location of the AGN and the peak of the diffuse emission
(coincident with Src~4), indicating that the most vigorous
star-formation is not spatially coincident with the AGN. 

This extended X-ray source (Src~4) with a luminosity of
$6.6\times10^{39}$~\ergs~ (0.1-10.0~keV band) is associated with the second brightest near IR peak in NGC3690 
(component B2). Its X-ray spectrum is
quite soft ($\Gamma\sim2.5$) but has large uncertainties.
It is also seen through a screen of relatively thick obscuring
material (\nh$\sim4.0\times10^{21}$~\cmsq), in agreement with the
extinction derived from NIR observations (AH00).  Its
intrinsic luminosity is $\sim2.5\times10^{40}$~\ergs~ similar to the
total luminosity from the central region of M82 (excluding the bright
ULX source), and consistent with this component of NGC~3690 being associated
with a compact starburst region.

Similar results are derived for the other sources with available X-ray spectra.
The relatively large column
density measured towards them  indicates that most of
the star-formation in Arp~299 is enshrouded by dust,
in agreement with results from the mid-IR band (Charmandaris \etal
2002), and as expected for a young burst of star-formation.
Stellar population evolutionary synthesis models give ages of                                                                                                 
$\sim5-15$~Myrs (AH00), for most star-forming regions, indicating that the
dominant X-ray components will be High Mass X-ray binaries and
supernova remnants.

Finally, the non detection of X-ray emission from component C$'$ is
quite consistent with a very young region (4~Myr) undergoing an instantaneous burst of
star-formation (AH00). A  calculation similar to that made above for  IC~694, gives a total
number of 60~HMXBs from the  Ly$_{c}$ emission of C$'$. 
However, given its probable young age, very few stars in binary systems have produced
compact objects and even fewer will have reached the end of their main
sequence lifetimes,  when they become observable as X-ray  binaries. 
Therefore the expected X-ray luminosity of this component 
would be much less than $\sim10^{39}$~\ergs~ which is the
typical detection limit of our observation.


\section{Conclusions}

We present preliminary results from a short \chandra ACIS-I
observation of the merger system Arp~299. This observation
reveals that the X-ray emission is produced by a population of
discrete sources which dominate above 2.0~keV, as well as emission
from a diffuse softer component.  These properties are very similar to
those found for other well studied merging star-forming galaxies
(e.g. Antennae: Fabbiano \etal 2001, Zezas \etal 2002; NGC~3256: Lira
\etal 2002). 
 However, there is a single very hard source in the nuclear region of 
 NGC~3690 which produces about 15\% of the observed
galactic emission above 2.0~keV in the \chandra energy band, which we identify with the AGN
reported by Della Ceca \etal (2002). The X-ray and multiwavelength properties
of the other near infrared nuclei are fully consistent with those found 
in other galaxies with nuclear starbursts, although 
the possibility of a relativelty weak mildly obscured AGN cannot be excluded in the
case of Src16 (assocaited with near infrared source A).

These results together with the recent detection of two AGNs
coincident which the double nuclei of the LIRG NGC~6240 (Komossa \etal
2002), suggest that galaxy mergers  are a very efficient means of
fueling AGNs (e.g. Combes \etal 2001). However, at the same time
mergers may result in burying AGN within large
concentrations of obscuring material, thereby rendering them invisible in all but the high
energy X-ray domain. Although the AGN ouput may not dominate the bolometric energy budget,
nevertheless their detection is important for any complete AGN demographic 
study, and also in furthering our understanding of the AGN/starburst connection.

\acknowledgments

We thank Susan Neff for providing results from the radio observations, 
prior to publication.
 We thank the CXC DS and SDS teams for the pipeline reduction the data and 
developing the software used for the pipeline reduction (SDP) and subsequent analysis
(CIAO). 
This work has been partly supported by NASA Grant  G01-2116X. 

{}

\makeatletter
\def\jnl@aj{AJ}
\ifx\revtex@jnl\jnl@aj\let\tablebreak=\nl\fi
\makeatother

\clearpage

\begin{deluxetable}{lccccccccc}
\tabletypesize{\tiny}
\tablecolumns{10}
\tablewidth{0pt}
\tablecaption{Properties of the discrete sources}
\tablehead{ 
\colhead{Src} & \colhead{RA (J2000)} & \colhead{Dec (J2000)} &
\colhead{Net} & \colhead{Sign.} & \colhead{HR1} & \colhead{HR2} &
\colhead{HR3} & \colhead{$\rm{L_{X}^{obs}}$, ($\rm{L_{X}^{corr}}$)} & \colhead{Notes$^{1}$} \\
\colhead{} & \colhead{h m s} & \colhead{\deg $'$ $''$} & \colhead{counts} & \colhead{$\sigma$} & \colhead{(S-M)/(S+M)} & \colhead{(S-H)/(S+H)} &
\colhead{(M-H)/(M+H)} & \colhead{$10^{39}$~\ergs}} 
\startdata
1 & 11 28 26.8 & +58 34 07.0  & 20.3  & 9.3  & $ -0.48 \pm 0.35$ & $ 0.17 \pm 0.25$ & $ -0.33 \pm 0.36  $ &  1.7 (2.0) \\

2 & 11 28 30.7 & +58 33 49.0  & 233.3 & 27.8 & $ 0.47 \pm 0.09 $ & $0.37 \pm 0.13 $ & $ 0.71 \pm 0.12 $ & 20.1 (22.5) 
& Ext; NIR: C1 \\
3 & 11 28 30.7 & +58 33 44.5  & 68.6  & 11.2 & $ 0.13 \pm 0.15 $ & $-0.03 \pm 0.15$ & $ 0.10 \pm 0.15 $ & 5.9 (6.6) \\

4 & 11 28 30.8 & +58 33 42.6  & 105.5 & 15.2 & $ 0.37 \pm 0.13 $ & $0.02 \pm 0.14 $ & $ 0.39 \pm 0.13 $ & 9.1 (10.2) 
& Ext; NIR: B2 \\
5 & 11 28 30.9 & +58 33 45.0  & 248.3 & 41.0 & $ 0.32 \pm 0.09 $ & $-0.26 \pm 0.09$ & $ 0.07 \pm 0.07 $ & 21.4 (23.9) 
& NIR: B21 \\
6 & 11 28 31.0 & +58 33 41.2  & 74.8  & 10.6 & $ 0.20 \pm 0.17 $ & $-0.44 \pm 0.17$ & $ -0.26 \pm 0.14$ & 6.4 (7.2) 
& Ext; NIR: B1 \\
7 & 11 28 31.2 & +58 33 26.2  & 12.6  & 4.6  & $ 0.37 \pm 0.35 $ & $0.57 \pm 0.65 $ & $ 0.78 \pm 0.56 $ & 1.1 (1.2) \\

8 & 11 28 31.7 & +58 33 49.5  & 38.4  & 8.8  & $ -0.29 \pm 0.20$ & $0.52 \pm 0.26 $ & $ 0.26 \pm 0.26 $ & 3.3 (3.7) 
& Ext; Rad: C$''$\\
9 & 11 28 32.3 & +58 33 18.4  & 52.3  & 23.6 & $ 0.07 \pm 0.16 $ & $0.13 \pm 0.18 $ & $ 0.19 \pm 0.18 $ & 4.5 (5.0) \\

10 & 11 28 33.1 & +58 33 37.0 & 29.3  & 7.2  & $ 0.49 \pm 0.30 $ & $-0.38 \pm 0.30$ & $ 0.13 \pm 0.21 $ & 2.5 (2.8) 
& Rad.: IC~694-1990\\
11 & 11 28 33.1 & +58 33 56.0 & 19.3  & 4.9  & $ 0.24 \pm 0.34 $ & $-0.40 \pm 0.34$ & $ -0.17 \pm 0.27$ & 1.7 (1.9) \\

12 & 11 28 33.2 & +58 33 43.9 & 36.1  & 6.8  & $ 0.28 \pm 0.21 $ & $0.18 \pm 0.25 $ & $ 0.44 \pm 0.25 $ & 3.1 (3.5) \\

13 & 11 28 33.3 & +58 34 02.9 & 46.6  & 15.4 & $ -0.44 \pm 0.22$ & $0.27 \pm 0.18 $ & $ -0.19 \pm 0.22$ & 4.0 (4.5) \\

14 & 11 28 33.4 & +58 33 46.9 & 60.4  & 11.7 & $ 0.49 \pm 0.19 $ & $-0.11 \pm 0.20$ & $ 0.40 \pm 0.18 $ & 5.2 (5.8) \\

15 & 11 28 33.7 & +58 33 51.2 & 42.2  & 9.0  & $ 0.78 \pm 0.27 $ & $0.38 \pm 0.47 $ & $ 0.89 \pm 0.30 $ & 3.6 (4.1) \\

16 & 11 28 33.7 & +58 33 47.2 & 132.6 & 15.6 & $ -0.22 \pm 0.14$ & $-0.36 \pm 0.11$ & $ -0.53 \pm 0.14$ & 11.4 (12.8) 
& Ext; NIR: A\\
17 & 11 28 34.1 & +58 33 40.0 & 30.0  & 6.4  & $ -0.48 \pm 0.29$ & $0.19 \pm 0.21 $ & $ -0.32 \pm 0.29$ & 2.6 (2.9) \\

18 & 11 28 37.5 & +58 33 40.9 & 9.2   & 4.1  & $ 0.15 \pm 0.34 $ & $1.08 \pm 1.08 $ & $ 1.06 \pm 0.91 $ & 0.8 (0.9) \\
\enddata
\tablenotetext{1}{Column~1: Source identifier; Column~2: RA (J2000);
Column~3: Dec (J2000); Column~4: Net number of counts in 
the 0.5-8.0~keV band;  Column~5: Significance of the source
with respect to the local background based on Gehrels statistic
(Gehrels 1986); Columns 6, 7 and 8:  Hardness ratios defined as
HR1=(S-M)/(S+M), HR2=(M-H)/(M+H), HR3=(S-H)/(S+H) where S, M and H
correspond to the numbers of net counts in  the (0.5-1.5)~keV, (1.5-2.5)~keV and (2.5-7.0)~keV 
bands respectively. Column~9: observed and absorption corrected luminosity of each source (0.1-10.0~keV) assuming a $\Gamma=1.7$ power-law absorbed by the
Galactic line-of-sight column density (\nh=$1.0\times10^{20}$\cmsq;
Stark \etal 1992); Column~10:  Near-Infrared (NIR) and
radio associations of the sources. In the same Column we denote the
extended sources (based on {\textit{wavedetect}} results from the full 
band image).}
\end{deluxetable}


\makeatletter
\def\jnl@aj{AJ}
\ifx\revtex@jnl\jnl@aj\let\tablebreak=\nl\fi
\makeatother

\begin{deluxetable}{lccccccccc}
\tabletypesize{\scriptsize}
\tablecolumns{10}
\tablewidth{0pt}
\tablecaption{Spectral fitting results}
\tablehead{ \colhead{Src} & \colhead{$\Gamma$} & \colhead{\nh} &
\colhead{\x2 (dof)} & \colhead{$\rm{L_{X}^{obs}}$, ($\rm{L_{X}^{corr}}$)$^{1}$} & \colhead{$\rm{L_{X}^{obs}}$, ($\rm{L_{X}^{corr}}$)$^{1}$}  \\
\colhead{} & \colhead{} & \colhead{$10^{22}$~\cmsq} & \colhead{} &
\colhead{(0.1-2.0)~keV} & \colhead{(2.0-10.0)~keV}} 
\startdata
2 &  $2.76^{+0.68}_{-0.47}$  & $0.29^{+0.18}_{-0.14}$  &  13.3 (10) & 7.4 (83.0) & 6.4 (6.6) \\
3 &  $1.35^{+1.4}_{-2.03}$   & $0.41  (<1.67)$         &  0.9 (1)   & 1.4 (4.4) & 9.2 (9.5) \\
4 &  $2.53^{+1.74}_{-1.04}$  & $0.41^{+0.60}_{-0.34}$  &  4.4 (3)   & 2.7 (22.1) & 3.9 (4.1) \\
6$^{2}$ &  $-2.46^{+0.32}_{-0.18}$ & $0.19^{+0.39}_{-0.16}$  & 8.8 (9) &  5.4 (17.1) & 66.3 (66.6) \\
  &  $1.89^{+1.61}_{-0.59}$  &   &   \\
14 & $2.57^{+2.88}_{-1.08}$  & $0.45^{+0.95}_{-0.44}$  &  1.2 (1) & 1.7 (19.9) & 2.5 (2.7) \\
16 & $1.46^{+0.85}_{-0.64}$  & $1.19^{+1.02}_{-0.74}$  &  1.6 (6) & 1.8 (13.6) & 2.1 (2.3) \\
\enddata
\tablenotetext{1}{Observed and absorption corrected luminosity in
units of $10^{39}$~\ergs.}
\tablenotetext{2}{The spectral parameters for this source are for a double
power-law model absorbed by a single column density.}
\end{deluxetable}

\clearpage

\begin{figure}
\includegraphics[width=15.0cm]{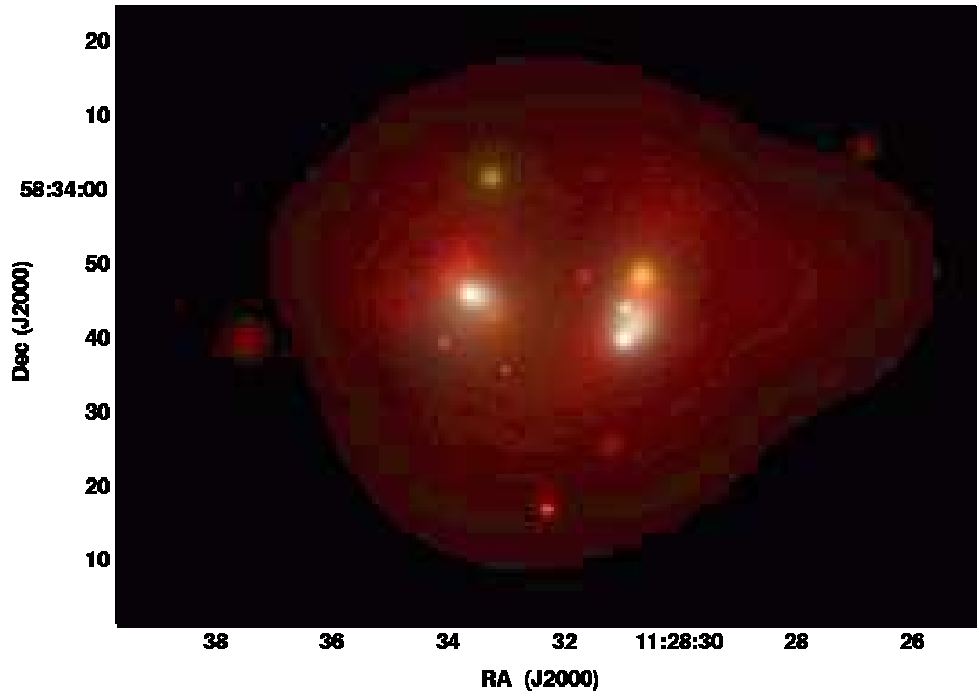}
\caption{A true colour X-ray image of Arp~299. Red corresponds to the
soft band (0.5-2.5~keV), green to the medium band (2.5-4.0~keV) and
blue to the hard band (4.0-8.0~keV). The size of the image is $2'$
(North is up and East is left).}
\end{figure}

\clearpage

\begin{figure}
\caption{A full band  (0.5-7.0~keV) adaptively
smoothed image of Arp~299 together with the discrete sources following
the notation of Table~1. The source regions correspond to the
appertures used the estimate their intensities and extract their spectra.}
\end{figure}

\end{document}